\documentclass{SciPost}

\hypersetup{
    colorlinks,
    linkcolor={red!50!black},
    citecolor={blue!50!black},
    urlcolor={blue!80!black}
}

\usepackage[bitstream-charter]{mathdesign}
\usepackage{float}

\DeclareSymbolFont{usualmathcal}{OMS}{cmsy}{m}{n}
\DeclareSymbolFontAlphabet{\mathcal}{usualmathcal}

\fancypagestyle{SPstyle}{
\fancyhf{}
\lhead{\colorbox{scipostdeepblue}{\bf \color{white} ~SciPost Physics Proceedings }}
\rhead{{\bf \color{scipostdeepblue} ~Submission }}

\fancyfoot[C]{\textbf{\thepage}}
}
\usepackage{enumitem} 

\usepackage{booktabs}
\usepackage{amsmath}
\usepackage{tabularx}
\usepackage{xcolor}
\usepackage{colortbl}
\definecolor{lightblue}{RGB}{173,216,230}
\definecolor{pink}{RGB}{255,182,193}
\definecolor{lightpink}{RGB}{255,204,204}

\begin{document}

\pagestyle{SPstyle}

\begin{center}{\Large \textbf{\color{scipostdeepblue}{
Quark-Gluon tagging performance at the High-Luminosity LHC using constituent-based transformer models \\
}}}\end{center}

\begin{center}\textbf{
Florencia L. Castillo\textsuperscript{1$\star$} and Jessica Lev\^eque
\textsuperscript{1} 
on behalf of the ATLAS Collaboration 
}\end{center}

\begin{center}
{\bf 1} Laboratoire d’Annecy de Physique des Particules (LAPP), Université Savoie Mont Blanc, CNRS/IN2P3, Annecy, France\\
$\star$ \href{mailto:email1}{\small fcastill@cern.ch}\,
\end{center}

\definecolor{palegray}{gray}{0.95}
\begin{center}
\colorbox{palegray}{
  \begin{tabular}{rr}
  \begin{minipage}{0.37\textwidth}
    \includegraphics[width=60mm]{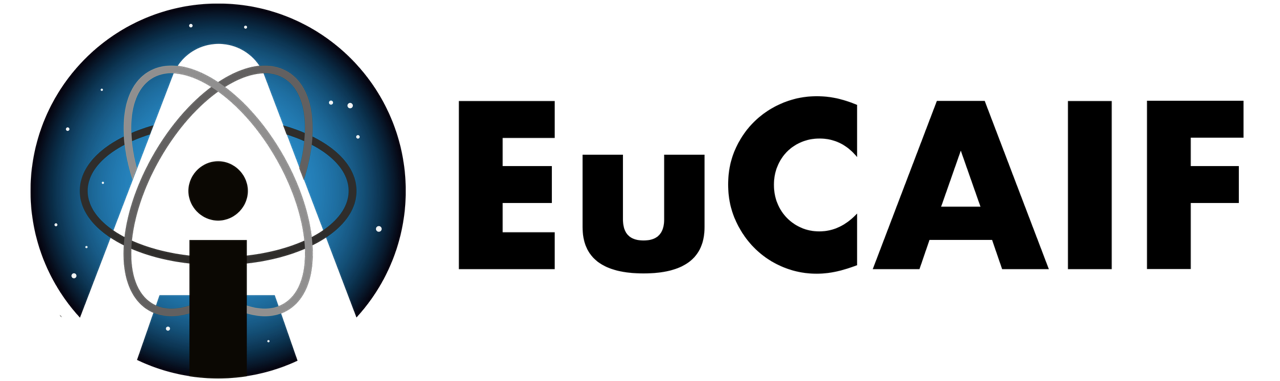}
  \end{minipage}
  &
  \begin{minipage}{0.5\textwidth}
    \vspace{5pt}
    \vspace{0.5\baselineskip} 
    \begin{center} \hspace{5pt}
    {\it The 2nd European AI for Fundamental \\Physics Conference (EuCAIFCon2025)} \\
    {\it Cagliari, Sardinia, 16-20 June 2025
    }
    \vspace{0.5\baselineskip} 
    \vspace{5pt}
    \end{center}
    
  \end{minipage}
\end{tabular}
}
\end{center}

\section*{\color{scipostdeepblue}{Abstract}}
\textbf{\boldmath{%
Jet constituents provide a more detailed description of a jet’s radiation pattern than global observables. In simulations for ATLAS Run-2 data (2015-2018), transformer-based taggers trained on low-level inputs outperformed traditional methods using high-level variables with conventional neural networks for quark–gluon discrimination. With the upcoming High-Luminosity LHC (HL-LHC), which will deliver higher luminosity and energy, the ATLAS detector will be upgraded with an extended Inner Tracker covering the forward region, previously uncovered by a tracking detector. This work studies how these upgrades will improve the accuracy and robustness of quark–gluon jet taggers.
}}

\vspace{\baselineskip}

\noindent\textcolor{white!90!black}{%
\fbox{\parbox{0.975\linewidth}{%
\textcolor{white!40!black}{\begin{tabular}{lr}%
  \begin{minipage}{0.6\textwidth}%
    {\small Copyright attribution to authors. \newline
    This work is a submission to SciPost Phys. Proc. \newline
    License information to appear upon publication. \newline
    Publication information to appear upon publication.}
  \end{minipage} & \begin{minipage}{0.4\textwidth}
    {\small Received Date \newline Accepted Date \newline Published Date}%
  \end{minipage}
\end{tabular}}
}}
}


\vspace{10pt}
\noindent\rule{\textwidth}{1pt}
\tableofcontents
\noindent\rule{\textwidth}{1pt}
\vspace{10pt}

{\tiny{Copyright 2025 CERN for the benefit of the ATLAS Collaboration. Reproduction of this article or parts of it is allowed as specified in the CC-BY-4.0 license.}}


\section{Introduction}
\label{sec:intro}
Quark–gluon tagging distinguishes narrower, harder quark-initiated jets from broader, softer gluon-initiated jets, which is crucial for enhancing signal–background separation in processes such as Vector Boson Fusion (VBF) and Vector Boson Scattering (VBS), where forward jets ($|y| > 2.5$) play a key role, and also provides benefits in searches for supersymmetry (SUSY) and heavy resonances \cite{ATL-PHYS-PUB-2018-023}.

The High-Luminosity LHC (HL-LHC), starting in 2030, will provide up to 3000 fb$^{-1}$ of data under challenging conditions, with an average of 140 pile-up interactions per bunch crossing. To address this environment, the ATLAS detector is going to be upgraded with an all-silicon Inner Tracker (ITk), extending charged-particle tracking to the forward region \cite{{ATLAS:2024ITk}}. This study investigates how these detector enhancements, combined with transformer-based models like the Particle Transformer (ParT), affect quark--gluon tagging using low-level jet data. In particular, we assess whether the performance observed in Run-2 simulations \cite{ATL-PHYS-PUB-2023-032} is maintained under HL-LHC conditions and how much forward-tracking information further improves the tagger compared to two fully connected (FC) baselines: one that employs eight high-level jet variables optimized for jet characterization, and an FC-reduced version that emulates the ATLAS quark–gluon tagger using five high-level variables from Run~2 analyses~\cite{ATLAS:2023dyu}.


\vspace{-0.5cm}

\section{Methodology}
\label{sec:method}
Taggers are trained on simulated VBF Higgs samples (Powheg\cite{Nason:2009ai}+Herwig7\cite{Bellm:2015jjp}) and dijet samples (Pythia8 \cite{Sjostrand:2014zea}) under HL-LHC conditions with an average pile-up of 140 interactions per bunch crossing. Jets are reconstructed using the anti-\(k_t\) algorithm (\(R=0.4\)) from Particle Flow Objects (PFOs), combining calorimeter topo-clusters and matched tracks, these are referred as jet constituents. The jet transverse momentum (\(p_T\)) spectrum is flattened during training, with uniform weights applied for evaluation. Two leading jets with \(p_T > 20\) GeV are selected in two regions: central (\(|y| < 2.5\)) and forward (\(2.5 < |y| < 4.0\)). The tagger descriptions and input variables are summarized in Table~\ref{tab:taggers_alt}.


\begin{table}[h]
\centering
\small
\begin{tabularx}{\textwidth}{>{\raggedright\arraybackslash}p{2.2cm} >{\raggedright\arraybackslash}X >{\raggedright\arraybackslash}X}
\toprule
\rowcolor{lightgray}
\textbf{Tagger} & \textbf{Description} & \textbf{Features} \\
\midrule
\rowcolor{lightblue}
\textbf{ParT} & Processes up to 50 PFOs per jet, ordered by descending \(p_T\). Concatenates topo-tower, track, and constituent inputs in the forward region for HL-LHC, extending ATLAS Run~2 configurations \cite{ATL-PHYS-PUB-2023-032}. & \textbf{Single-constituent}: Relative rapidity (\(\Delta y^a = y^a - y^{\text{jet}}\)), azimuthal angle difference (\(\Delta \phi^a = \phi^a - \phi^{\text{jet}}\)), \(\Delta R^a = \sqrt{(\Delta y^a)^2 + (\Delta \phi^a)^2}\), \(\log p_T^a\), \(\log E^a\), \(\log(p_T^a / p_T^{\text{jet}})\), \(\log(E^a / E^{\text{jet}})\), constituent mass (\(m^a\)). \\
\rowcolor{lightblue}
 & & \textbf{Pairwise}: Angular separation (\(\Delta R_{ab} = \sqrt{(y^a - y^b)^2 + (\phi^a - \phi^b)^2}\)), invariant mass (\(m_{ab}^2 = (p^{\mu,a} + p^{\mu,b})^2\)), Lund splitting variables (\(k_T = \min(p_T^a, p_T^b) \cdot \Delta R_{ab}\), \(z = \min(p_T^a, p_T^b) / (p_T^a + p_T^b)\)). \\
\midrule
\rowcolor{pink}
\textbf{FC} & Employs eight high-level jet variables for tagging, optimized for jet characterization. & Jet transverse momentum ($p_T$), jet mass ($m$), electromagnetic fraction (EMFrac), jet width (from PFOs, charged PFOs, and tracks with $p_T > 1\,\text{GeV}$), number of PFOs, and number of charged PFOs ($p_T > 1\,\text{GeV}$) \\
\midrule
\rowcolor{lightpink}
\textbf{FC-reduced} & Emulates ATLAS quark--gluon tagging with five high-level variables from Run~2 analyses\cite{ATLAS:2023dyu}. & Jet \(p_T\), pseudorapidity (\(\eta\)), number of PFOs, PFO width (\(w^{\text{PFO}}\)), two-point energy correlation (\(C_1^{\beta=0.2}\)). \\
\bottomrule
\end{tabularx}
\caption{Characteristics and features of taggers. Indices \(a, b\) denote different PFOs within a jet.}
\label{tab:taggers_alt}
\end{table}

\vspace{-0.5cm}
\section{Results}
\label{sec:results}
\vspace{-0.2cm}

The tagger performance is quantified using gluon-jet rejection ($\epsilon_g^{-1}$) at a fixed quark efficiency of $\epsilon_q = 0.5$. It is evaluated as a function of jet rapidity ($|y|$) in low- and high-$p_T$ ranges, with results shown for the central (Sec.~\ref{sub:central}) and forward (Sec.~\ref{sub:forward}) regions. Pile-up robustness is studied for 60, 140, and 200 additional interactions per bunch crossing (Sec.~\ref{sub:pu}).

\subsection{Central Region}\label{sub:central}
\begin{figure}[!h]
\centering
    \includegraphics[width=0.45\textwidth]{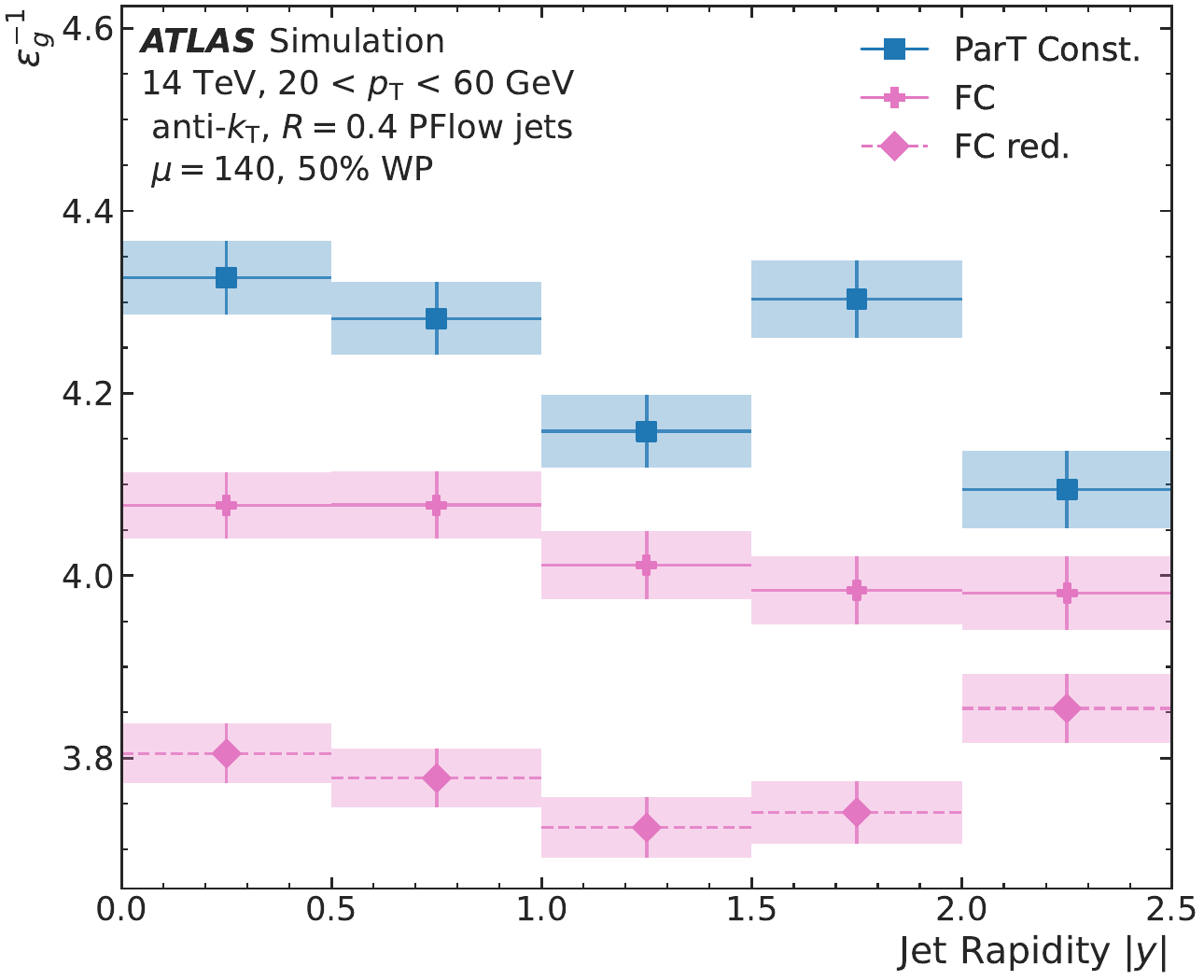}
    \includegraphics[width=0.45\textwidth]{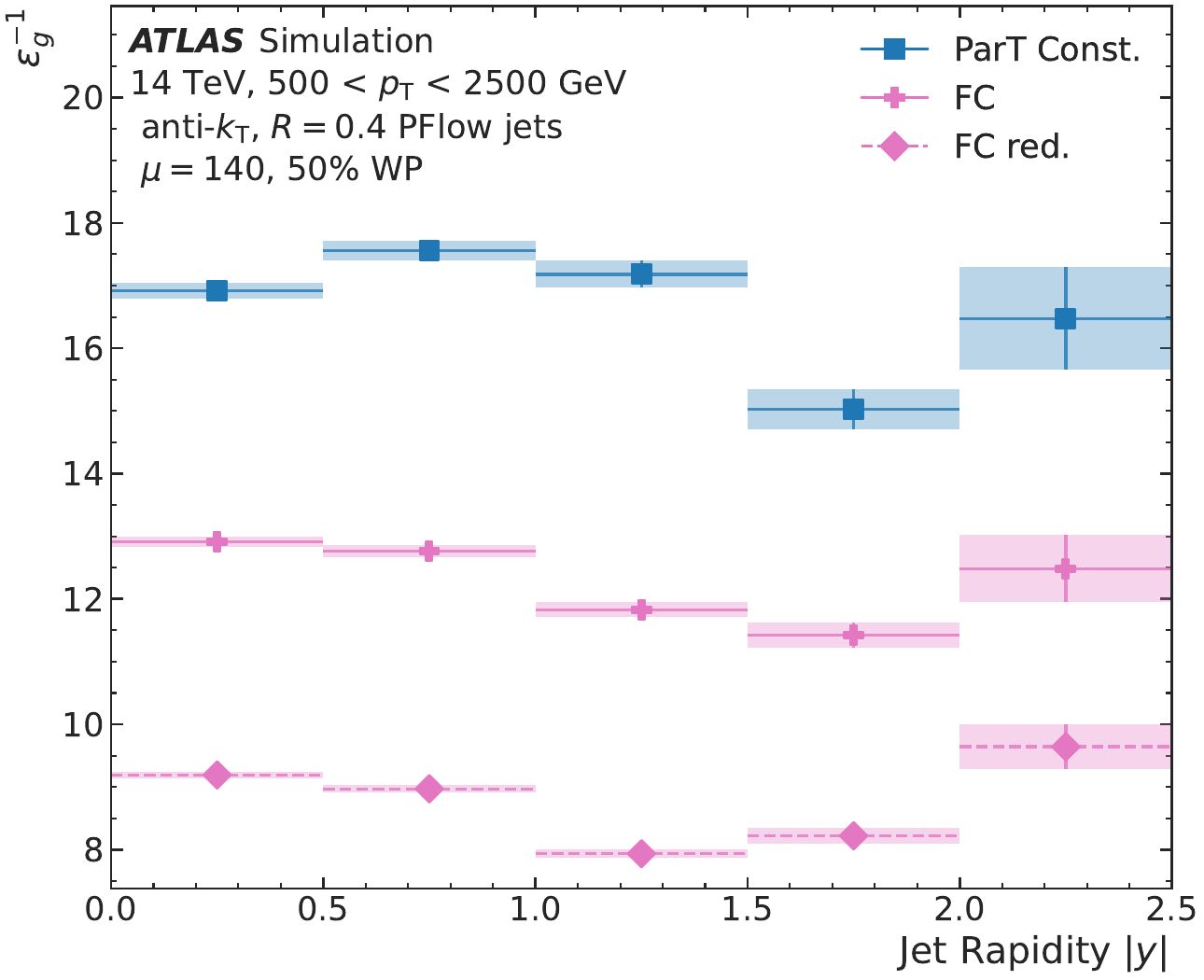}
    \caption{Gluon jet rejection (\(\epsilon_g^{-1}\)) vs. jet rapidity (\(|y|\)) in the central region, across low- and high-\(p_T\) at \(\epsilon_q = 0.5\). ParT (blue) and FC taggers (solid and dashed pink) are compared under HL-LHC conditions with pile-up 140. Error bars show statistical uncertainties  \cite{ATLAS:JETM-2025-01}.}
    \label{fig:central}
\end{figure}

In the central region, across both low- and high-\(p_T\) ranges (Figures \ref{fig:central}), the ParT tagger outperforms the FC tagger, achieving approximately 10\% better gluon rejection at low \(p_T\) and up to 25\% improvement at high \(p_T\), thanks to its detailed constituent-level inputs.

\subsection{Forward Region} \label{sub:forward}

\begin{figure}[!h]
\centering
    \includegraphics[width=0.45\textwidth]{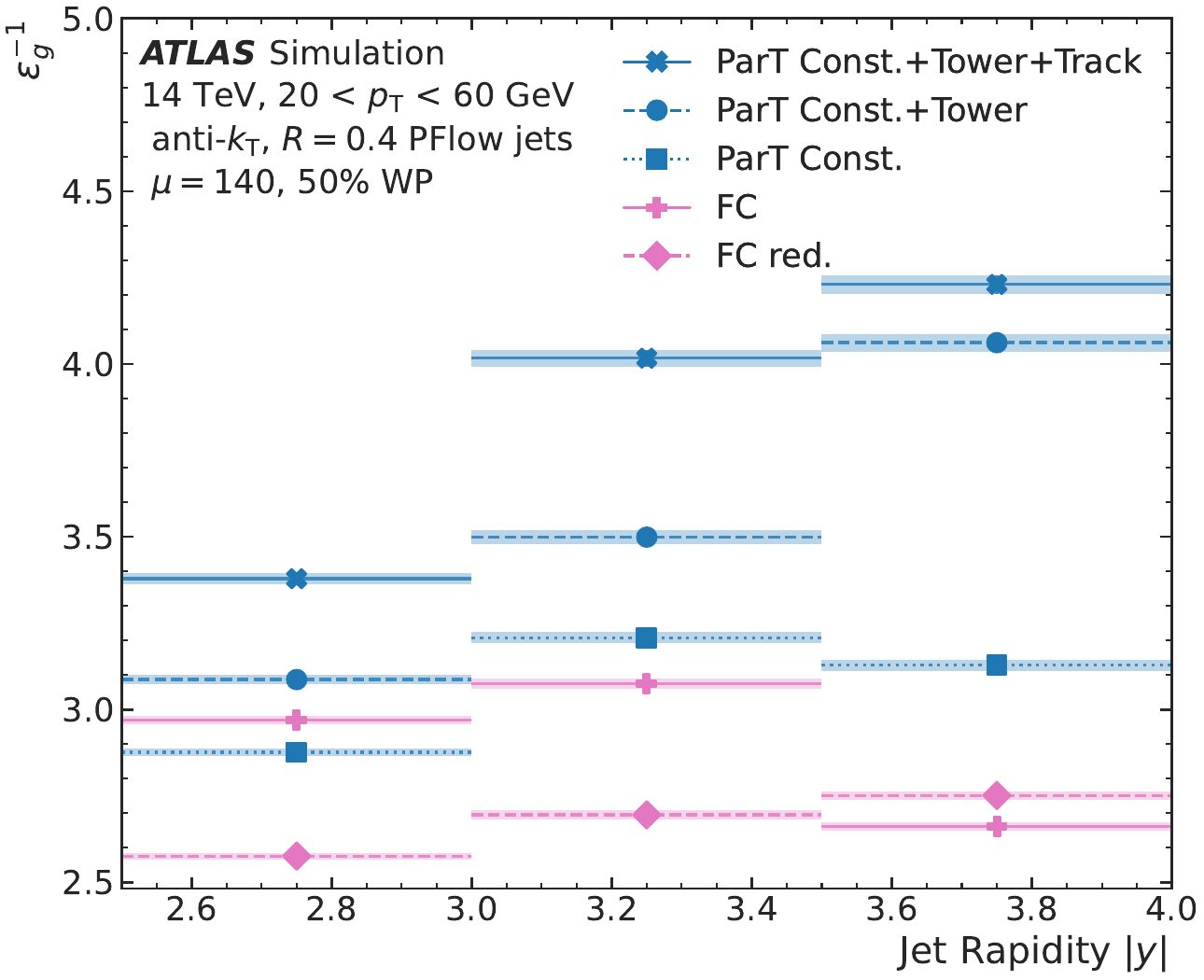}
    \includegraphics[width=0.45\textwidth]{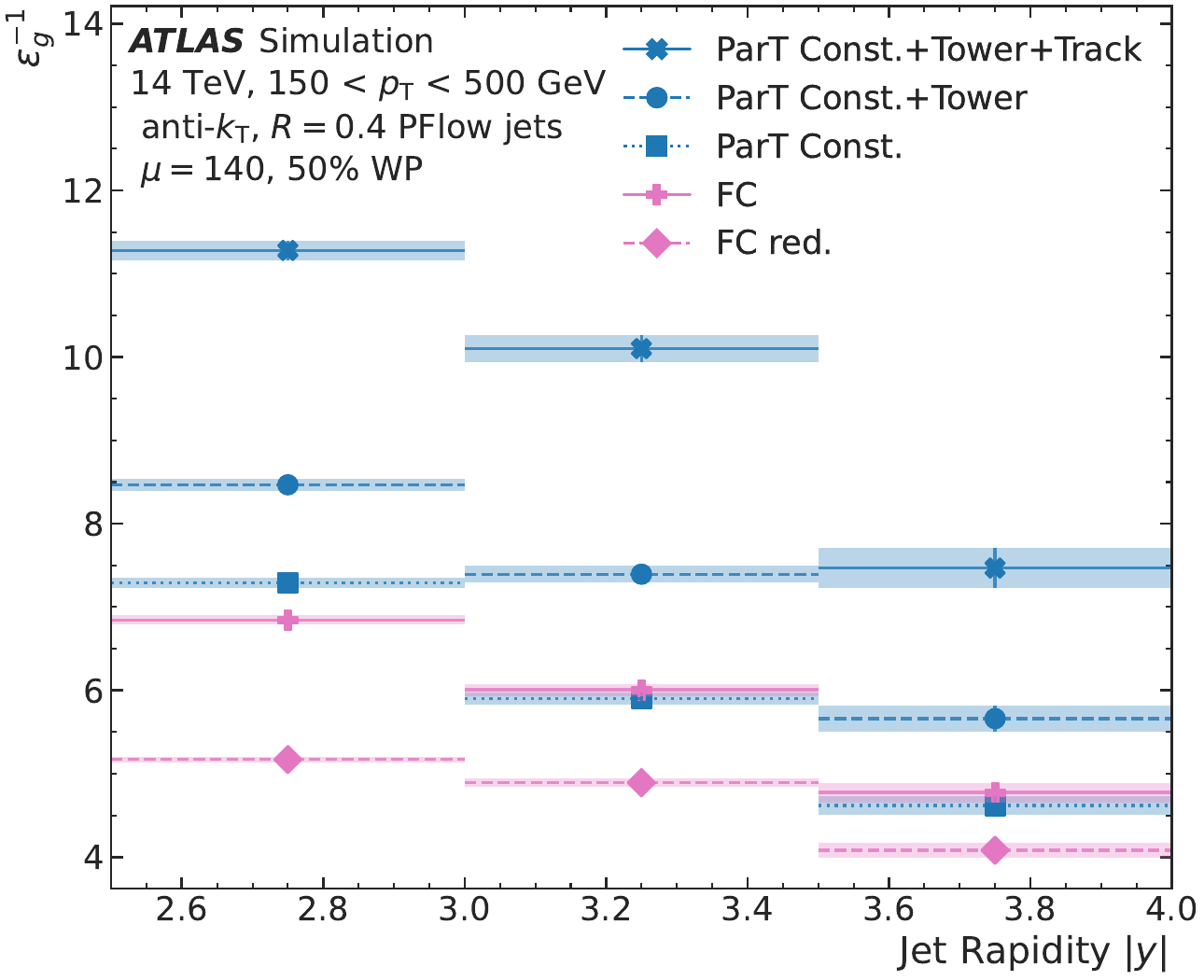}
    \caption{Gluon jet rejection (\(\epsilon_g^{-1}\)) vs. jet rapidity (\(|y|\)) in the forward region, low-\(p_T\) and range, at \(\epsilon_q = 0.5\). ParT (blue) and FC taggers (solid and dashed pink) are compared under HL-LHC conditions with pile-up 140. Error bars show statistical uncertainties \cite{ATLAS:JETM-2025-01}}
    \label{fig:forward}
\end{figure}

In the forward region (Figures \ref{fig:forward}), the ParT tagger achieves 20--30\% better gluon rejection than the FC and FC-reduced taggers by using constituent, track, and topo-tower information. With only constituents (ParT Const.), performance is reduced due to the drop in track efficiency \cite{ATLAS:2024ITk}, limiting the effectiveness of track-dependent constituent features. Adding tracks and topo-towers (ParT Const. + Tower + Track) improves performance, as available tracks provide additional discriminating information and the transformer can handle missing tracks. Topo-towers alone (ParT Const. + Tower) offer complementary gains.

\subsection{Pile-up robustness}\label{sub:pu}
\begin{figure}[!h]
    \centering
    \includegraphics[width=0.45\textwidth]{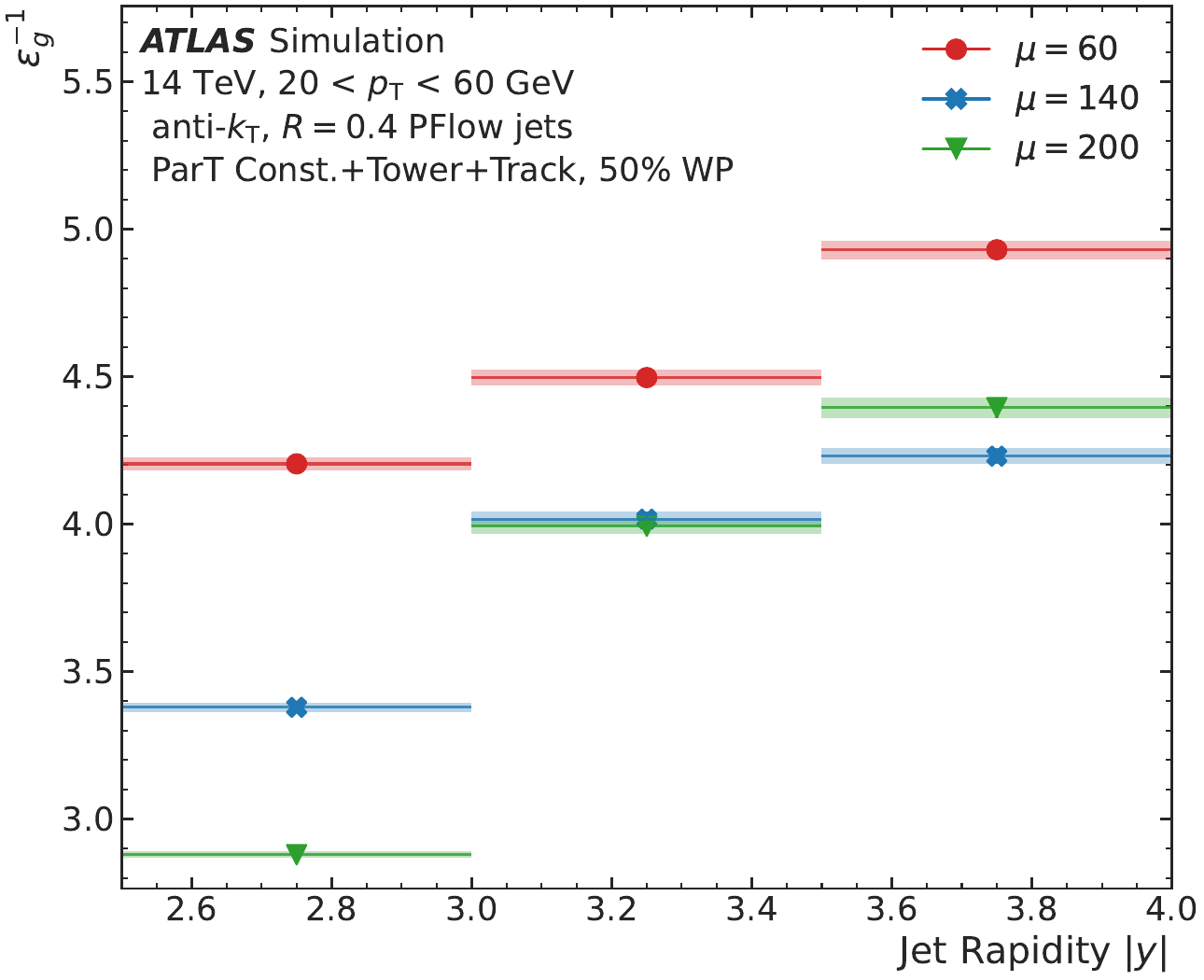}
    \includegraphics[width=0.45\textwidth]{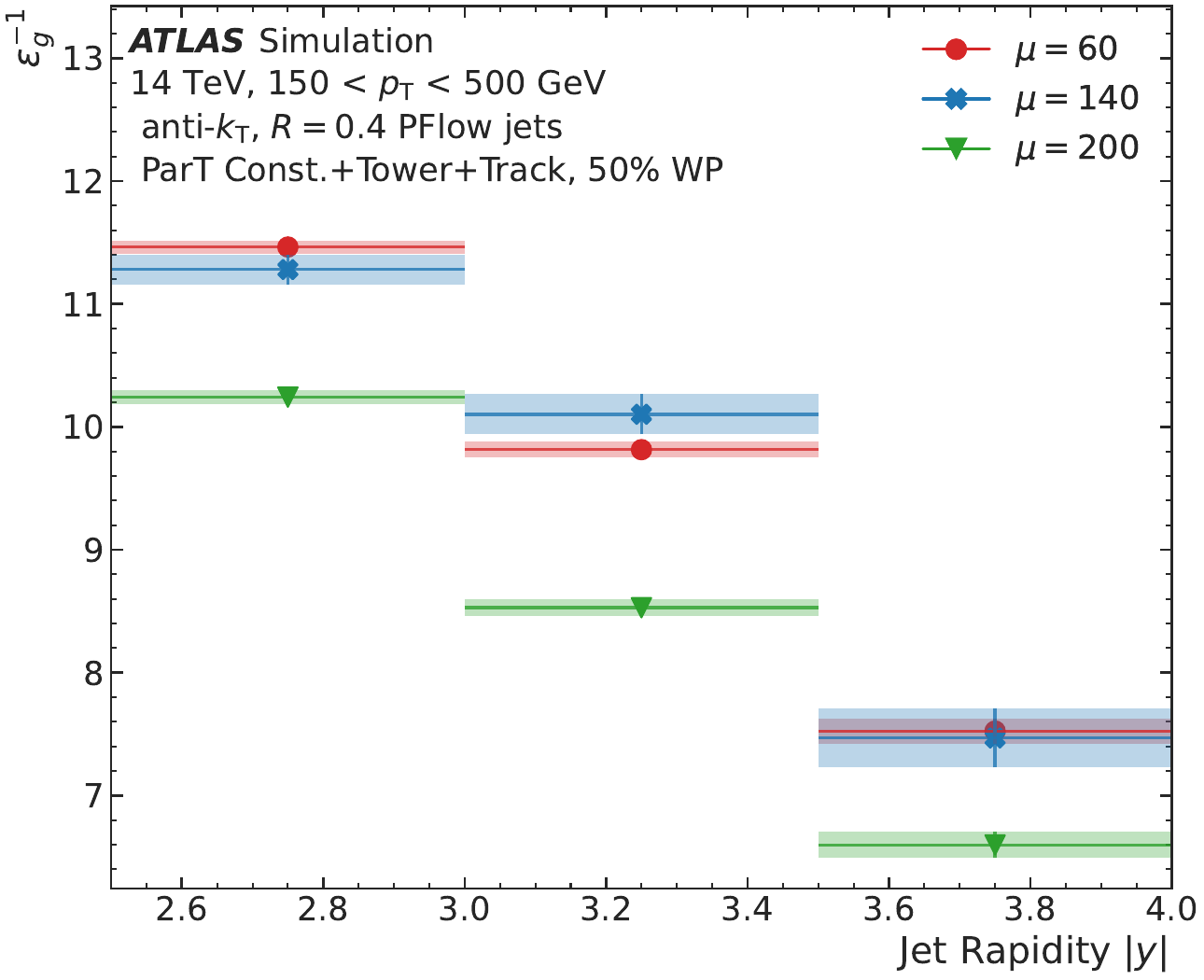}
    \caption{Gluon jet rejection (\(\epsilon_g^{-1}\)) as a function of jet rapidity (\(|y|\)) in the central region for low- and high-\(p_T\) jets at \(\epsilon_q = 0.5\). Results are shown for the ParT tagger using concatenated constituent, topo-tower, and tracking inputs, evaluated under pile-up conditions of 60 (red), 140 (blue), and 200 (green). Error bars indicate statistical uncertainties \cite{ATLAS:JETM-2025-01}.}
    \label{fig:pu_low_pt}
\end{figure}

Figure \ref{fig:pu_low_pt} shows ParT’s performance in the forward region under pile-up levels of 60, 140, and 200. Performance remains stable, with minimal degradation at higher pile-up, highlighting robustness for HL-LHC conditions.

\section{Conclusion}

\label{sec:conclusion}
This study shows that transformer-based ParT taggers trained on low-level jet information improve quark--gluon discrimination at the HL-LHC. Including ITk tracking further enhances performance, particularly in the forward region, yielding up to 30\% higher gluon rejection compared to FC taggers. The approach is robust against pile-up and is expected to increase the sensitivity of analyses such as VBF, VBS, and SUSY searches.
\vspace{-0.5cm}

\section*{Acknowledgements}
\paragraph{Funding information}
Supported by the Agence Nationale de la Recherche (ANR) under the program "Advanced Tracking Algorithms for Particle Physics (ATRAPP)" (ANR-21-CE31-0022) and CERN via the ATLAS Collaboration.


\end{document}